\documentclass[a4paper,twoside,10pt]{article}

\input{jgrg21.sty}

\begin{document}

\pagestyle{fancy}
\fancyhead{}
  \fancyhead[RO,LE]{\thepage}
  \fancyhead[LO]{N. Yunes}                  
  \fancyhead[RE]{Gravitational Waves from Compact Binaries as Probes of the Universe}
\rfoot{}
\cfoot{}
\lfoot{}

\label{O28-02}    

\title{Gravitational Waves from Compact Binaries \\ as Probes of the Universe}

\author{Nicol\'as Yunes \footnote{Email address: nyunes@physics.montana.edu}$^{(a)}$}

\address{$^{(a)}$Department of Physics, Montana State University, Bozeman, MT 59718, United States.}

\abstract{
The future detection of gravitational wave forces us to consider the many ways in which astrophysics, gravitational wave theory and fundamental theory will interact. In this paper, I summarize some recent work done to develop such an interface. In particular, I concentrate on how non-vacuum astrophysical environments can modify the gravitational wave signal emitted by compact binary inspirals, and whether signatures from the former are detectable by current and future gravitational wave detectors. I also describe the interface between gravitational wave modeling and fundamental theory, focusing on the status of the parameterized post-Einsteinian framework (a general framework to detect deviations away from General Relativity in future gravitational wave data) and its current data analysis implementation.   
}

\section{Multi-Messenger Gravitational Wave Astrophysics}

The future detection of gravitational waves opens fertile ground to develop
a collaboration between astrophysics, gravitational wave theory and data analysis. 
The traditional way in which such interdisciplinary work was thought to happen was fairly
sequential: astrophysics would predict which sources gravitational wave theorist should
model; theorist would provide highly accurate models for such systems; and data analysts
would take such models and implement them in pipelines to maximize the probability of
detecting a signal. An accurate astrophysical system and waveform model are crucial
as the first detections will consist of signals that are completely buried in instrumental noise.  

Once a signal has been detected, feedback in this cycle can occur. That is, data analysts can
extract the best-fit parameters for the signal detected and, given enough detections, create
population models of the distribution of system parameters in the low-redshift universe. For 
example, given a sufficient number of compact binary signals, one could construct the distribution
of masses, mass ratios, spins, etc. This information could then be returned to the astrophysics 
community to see how it matches predicted population models. 

Another route in which feedback can occur is through the data placing  
constraints on General Relativity. Given a single detection, data analysts can determine the
probability that the foundational assumptions of General Relativity remain valid in strongly
gravitating, highly dynamical systems, where tests of Einstein's theory currently do not exist. 
The most stringent tests we have today come from binary pulsar observations (see e.g.~\citep{O28-02_Kramer:2005ez}), but the gravitational compactness of such bodies, the characteristic mass of the system divided by its orbital separation, remains minuscule. Compact binary coalescences can in principle sample horizon-scale compactnesses, thus allowing for the most stringent tests yet.  

The interaction of gravitational wave theory with astrophysics is a two-way street. One way to proceed is to focus on how the dynamical evolution of a gravitational system affects the astrophysical environment. An example of this is the evolution of a circumbinary accretion disk when its binary black hole nucleus coalesces. In this process, the binary emits $\sim 5-10\%$ its rest mass in gravitational waves, depending on the orbital characteristics of the system. A change in rest mass leads to a modification in the Newtonian potential that the circumbinary accretion disk is bounded to, forcing the latter to readjust and leading to shocks in the material that then emits electromagnetic radiation (see e.g.~\citep{O28-02_Bode:2009mt,O28-02_Bogdanovic:2010he}).  

A different way to proceed is to focus on how the astrophysical environment modifies the gravitational wave evolution. Let us revisit the example given above, ie.~that of a circumbinary accretion disk with a comparable-mass binary black hole in a quasi-circular inspiral. One could imagine that the mere presence of the accretion disk might modify the evolution of the binary, for example due to angular momentum transport by the disk or the disk's self-gravity. Surely, these and other effects will have an impact on the binary's inspiral rate, but are such effects measurable by future gravitational wave detectors? For the current example, the answer is likely no. That is, the effect of the accretion disk is probably too small to be detectable by LISA, because the comparable mass binary's inertia is too great relative to the effect of the disk. 

The effect of the astrophysical environment on the gravitational wave signal is not always negligible, however, and the prototypical system where the former matters is extreme mass-ratio inspirals. These systems consist of a solar-mass compact object spiraling into a supermassive black hole in a generic orbit. Due to the extreme mass-ratio, any small perturbation produced by a non-vacuum, astrophysical environment can have drastic effects on the small object's trajectory. Moreover, such systems can remain in the sensitivity band of low-frequency gravitational wave detectors for tens of years, depositing hundreds of thousands of radians of phase information in any given year. Thus, any small perturbation can accumulate over a year, leading to an observable effect. In this paper, I will describe a few papers~\citep{O28-02_Yunes:2010sm,O28-02_Kocsis:2011dr,O28-02_Yunes:2011ws} that have studied two possible astrophysical environment effects on extreme-mass ratio inspiral gravitational waves: the effect of accretion disks and the effect of massive perturbers. 

The interaction of gravitational wave theory and data analysis with fundamental theory can also proceed in two different ways. Until very recently, the standard approach was {\emph{top-down}}: one would pick a modified gravity theory, derive its equations of motion and solve them to find its predicted gravitational wave radiation. Once a prediction had been made, one could construct a template bank for this modified theory and then, given a gravitational wave detection consistent with General Relativity, constrain how large the parameters of the modified theory can be. Examples of this approach exist for Brans-Dicke theory~\citep{O28-02_Will:1994fb,O28-02_Will:2004xi,O28-02_Berti:2005qd,O28-02_Stavridis:2010zz,O28-02_Arun:2009pq,O28-02_Keppel:2010qu,O28-02_Yagi:2009zm}, dynamical Chern-Simons gravity~\citep{O28-02_Sopuerta:2009iy,O28-02_Yunes:2009hc}, phenomenological massive graviton propagation~\citep{O28-02_Will:1997bb,O28-02_Scharre:2001hn,O28-02_Will:2004xi,O28-02_Berti:2005qd,O28-02_Yagi:2009zm,O28-02_Berti:2011jz}, gravitational Lorentz-violation~\citep{O28-02_Mirshekari:2011yq}, and gravitational parity violation~\citep{O28-02_Yunes:2010yf}, violations of Local Position Invariance~\citep{O28-02_Yunes:2009bv} and the existence of extra-dimensions~\citep{O28-02_Yagi:2011yu}.

The problem with any of these approaches is that the construction of template banks and subsequent analysis of detected signals is computationally expensive for each alternative theory. Moreover, it is unclear which modified theory effect to concentrate on, since none of the above models is particularly more theoretically appealing than General Relativity. For example, in the massive graviton case, there is no self-consistent theory at the level of the action that would lead to such massive graviton effects. 

Lacking such a particular compelling model, a new framework to constrain deviations from General Relativity was recently proposed: the parameterized post-Einsteinian (ppE) framework~\cite{O28-02_Yunes:2009ke}. This approach proposes to enhance the General Relativity template banks (spanned by system parameters) through the addition of new, well-motivated theory parameters. These parameters are such that when they acquire certain numerical values, they lead to waveforms predicted by all the modified gravity theories described above. The advantage of such an approach is that one need only concentrate on the construction of a single template bank. Matched filtering with elements from this bank would then allow the data to select the theory that best matches the signal, thus lifting the fundamental theoretical bias that General Relativity must be {\emph{a priori}} correct. In this paper, I will summarize the current status of the ppE framework. 

\section{Gravitational Wave Modeling}

Gravitational wave modeling has a long history and it would be quite pretentious and inappropriate to attempt a complete description here. Instead, I will only summarize the bits and pieces of gravitational wave modeling that will be necessary to understand the modeling of astrophysical and modified gravity effects, focusing on compact binary coalescences (inspiral-phase only) and ignoring any tidal deformation of the compact objects. I refer the interesting reader to the following review papers~\citep{O28-02_Blanchet:2006zz} for post-Newtonian, comparable-mass inspiral modeling,~\citep{O28-02_Barack:2009ux} for self-force, extreme-mass ratio modeling,~\citep{O28-02_2009GWN.....2....3Y} for approximate, extreme-mass ratio modeling, and references therein. 

\subsection{Comparable-Mass Compact Binary Quasi-Circular Inspirals}
\label{sec:O28-02_SPA}

Let us begin with the modeling of gravitational waves emitted by comparable-mass, compact binary inspirals. The observed quantity at the detector is not the time-domain waveform, but the response function: a quantity constructed from the contraction of the response tensor with the (transverse-traceless) gravitational wave metric perturbation: 
\begin{equation}
h(t) = F_+(\theta_S\,,\phi_S\,,\psi_S) h_+ + 
F_{\times}(\theta_S\,,\phi_S\,,\psi_S) h_{\times},
\end{equation}
where $h_{+,\times}$ are the plus- and cross-polarized gravitational wave metric perturbations, while $F_{+,\times}$ are beam-pattern functions . These functions characterize the response of the detector to an incoming gravitational wave and are slowly-varying.

The Fourier transform of the response function can be evaluated analytically if we assume that the phase varies on a much shorter time-scale than the amplitude, ie. in the stationary phase approximation~\citep{O28-02_Yunes:2009yz}. Let us define the cosine and sine responses for the dominant $\ell = 2$ mode
\begin{equation}
h(t) = h_C(t) + h_S(t)\,,
\qquad
h_C(t) = {\cal{A}} \; Q_C(\iota,\beta) \; \cos{2 \phi}\,,
\qquad
h_S(t) = {\cal{A}} \; Q_S(\iota,\beta) \; \sin{2 \phi}\,,
\end{equation}
where ${\cal A} = - {\cal{M}}/D_{L} \left(2 \pi {\cal{M}} F\right)^{2/3}$ is a slowly-varying amplitude, while
\begin{equation}
Q_C(\iota,\beta) 
\equiv 2 \left(1 + c_i^2\right) c_{2 \beta} F_+ - 4 c_i s_{2 \beta} F_{\times}\,,
\qquad
Q_S(\iota,\beta) \equiv 
2 \left(1 + c_i^2\right) s_{2 \beta} F_+ + 4 c_i c_{2 \beta} F_{\times}\,,
\end{equation}
and $c_{i} = \cos{\iota}$, $c_{2 \beta} = \cos{2 \beta}$ and $s_{2 \beta} = \sin{2 \beta}$, with $\iota$ and $\beta$ the inclination and polarization angles respectively. Then, the Fourier transform in the stationary phase approximation is 
\begin{equation}
\tilde{h}_{C}(f) = - \left(\frac{5}{384}\right)^{1/2} \pi^{-2/3} 
\frac{{\cal{M}}^{5/6}}{D_L} \; Q_C(\iota,\beta) \; 
\left[2 {F}(t_0)\right]^{-7/6} 
e^{-i \left(\Psi + \pi/4 \right)}\,,
\label{O28-02_SPA-FFT}
\end{equation}
while $\tilde{h}_{S}(f) = i \tilde{h}_{C \to S}(f)$. The phase of the gravitational wave Fourier transform is 
\begin{equation}
\Psi[F(t_{0})] = 2 \pi \int^{F(t_{0})} \frac{F'}{\dot{F}'} \left(2 - \frac{f}{F'}\right) dF'. 
\label{O28-02_Phase}
\end{equation}
The quantity $t_{0}$ is the {\emph{stationary-point}}, defined by $2 \dot{\phi}(t_{0}) = 2 \pi f$, essentially the region in which the Fourier transform does not average out. In Eq.~\eqref{O28-02_SPA-FFT}, $D_{L}$ is the luminosity distance, ${\cal{M}} = \eta^{3/5} m$ is the chirp mass, with $\eta = m_{1} m_{2}/m^{2}$ the symmetric mass ratio and $m = m_{1} + m_{2}$ the total mass, while $F$ is the orbital frequency. 

Gravitational wave detectors are the most sensitive to the phase of the response function, Eq.~\eqref{O28-02_Phase}, whose functional form is controlled by the orbital frequency's evolution equation. In vacuum General Relativity, the orbital frequency can only evolve due to the emission of gravitational waves. Using the balance law $\dot{E} = - \left<{\cal{L}}_{\rm GW} \right>$, i.e.~the rate of change of the binary's binding energy is exactly balanced by the flux of gravitational wave energy-momentum suitably averaged over several wavelength, one can derive
\begin{equation}
\frac{dF}{dt} \equiv \frac{\dot{E}}{dE/dF} \sim \frac{48}{5\pi {\cal M}^2} 
\left(2\pi {\cal M} F\right)^{11/3}
\end{equation}
to leading order in $2 \pi {\cal{M}} F \sim v^{3}$, where $v$ is the binary's orbital velocity. Of course, if Kepler's third law is modified or if additional sinks of energy-momentum are present, for example due to scalar-field emission or to accretion-disk induced angular momentum transport, then the gravitational wave phase will be modified accordingly. Ignoring these effects for the moment, the Fourier transform is then
\begin{equation}
\tilde{h}_{\rm (circ)} = - \left(\frac{5}{384}\right)^{1/2} \pi^{-2/3} 
\frac{{\cal{M}}^{5/6}}{D_L} \; Q(i,\beta) \; f^{-7/6}
\exp\left[
  {i \left(2 \pi f t_c  - \bar\phi_c - \frac{\pi}{4}  + \frac{3}{128} 
      \left( \pi {\cal{M}} f\right)^{-5/3} \right)} \right],
\end{equation}
where $Q = Q_{C} + i Q_{S}$ and we have retained only the Newtonian term in the phase. 

\subsection{Extreme Mass-ratio, Compact Binary Quasi-Circular Inspirals}
\label{sec:O28-02_EOB}

Let us now consider extreme-mass ratio inspirals in a quasi-circular orbit confined to the equatorial plane of the supermassive black hole. The dominant $\ell = 2$, gravitational wave mode accumulates a phase
\begin{equation}
\phi_{\rm GW}
=
2 \int^{t_f}_{t_f - T_{\rm obs}} \! \! \!  \! \! \!  \! \! \! \Omega(t) \; dt
= 2 \int^{R_f}_{R_0}  \! \! \!  \Omega(r) \frac{d r}{\dot{r}}\,
=
\frac{1}{16}\frac{M_{\rm SMBH}}{m_{\rm CO}} R_{f}^{5/2}
\left[\left(1 + \frac{\tau}{R_{f}^4} \right)^{5/8}-1\right]\,
\label{O28-02_accumulated-phase}
\end{equation}
where $T_{\rm obs}$ is the observation time, $\Omega$ is the orbital angular frequency, $M_{\rm SMBH}$ is the supermassive black hole mass, $m_{\rm CO}$ is the stellar-mass compact object's mass and $(R_{0},R_{f})$ are the $M_{\rm SMBH}$-normalized initial and final orbital radius respectively. The quantity $\tau$ here is the dimensionless observation time defined by
$\tau \equiv (256/{5}) (m_{\rm CO} T_{\rm obs}/{M_{\rm SMBH}^{2}})$. One can verify that for low-frequency gravitational wave detectors, with peak sensitivity in the micro- to deci-Hz band, the accumulated gravitational wave phase can reach millions of radians for separations in $(R_{0},R_{f}) \in (30 M_{\rm SMBH},R_{\rm ISCO})$, where $R_{\rm ISCO}$ is the radius of the innermost stable circular orbit for a test-particle on a geodesic of the background spacetime.

The accumulation of gravitational wave phase information clearly depends on the rate of inspiral, $\dot{r}$. This quantity can be computed from $\dot{r} = \dot{E}/(dE/dr)$, which again depends both on the binary's binding energy $E(r)$ and its rate of change. If additional sinks of energy are present in the problem or if the binding energy is deformed, then the accumulated gravitational wave phase will be different than what one would expect for a vacuum inspiral.  

The accumulated gravitational wave phase in Eq.~\eqref{O28-02_accumulated-phase} can be used as a quick measure of how important non-vacuum or non-General Relativity effects are, but this measure is only approximate since it uses Newtonian relations, which are highly inaccurate for extreme-mass ratio inspiral. A much more appropriate scheme is that of the {\emph{effective-one-body}} approach, recently implemented for extreme-mass ratio inspirals in~\citep{O28-02_Yunes:2009ef,O28-02_Yunes:2010zj} and references therein. In this scheme, one maps the two-body inspiral into an effective inspiral of a test-particle with mass $\eta$ in orbit around a supermassive black hole with mass $m$. The equations of motion then reduce to
the Hamilton-Jacobi equations 
\begin{align}
  \frac{dr}{d t} &= 
  \frac{A(r)}{\sqrt{D(r)}}\frac{\partial \widehat{H}^{\rm real}}
  {\partial p_{r_*}}\,, \quad \quad \frac{d \Phi}{d t} =
  \frac{\partial \widehat{H}^{\rm real}}  {\partial p_\Phi}\,,
  \label{eq:eobhamonetwo} \\
  \frac{d p_{r_*}}{d t} &=- 
  \frac{A(r)}{\sqrt{D(r)}}\,\frac{\partial \widehat{H}^{\rm real}}
  {\partial r}\,, \quad \quad \frac{d p_\Phi}{d t}=
  \widehat{\cal F}_\Phi\,.
  \label{O28-02_EOB-H-Eqs}
\end{align}
where $\widehat{H}^{\rm real} \equiv H^{\rm real}/\mu$ is the reduced (i.e., dimensionless) real Hamiltonian and $\widehat{\cal{F}}_{\Phi} := {{\cal{F}}}_{\Phi}/\mu$ is a reduced radiation-reaction force that controls the rate of inspiral (there is no radial component on average here because we are dealing with an equatorial, quasi-circular orbit). The azimuthal radiation-reaction force can be modeled via  $\widehat{\cal{F}}_{\Phi} = \eta^{-1} \dot{L} = - \eta^{-1} \dot{L}_{\rm GW} = - (\eta \Omega)^{-1} \dot{E}_{\rm GW}$, where $\dot{L}$ is the rate of change of the binary system's orbital angular momentum, while $\dot{L}_{\rm GW}$ and $\dot{E}_{\rm GW}$ are the angular momentum and energy radiated in gravitational waves. 

The energy flux is composed of several contributions. The most accurate model for the average, energy flux radiated to infinity is the factorized resummation 
\begin{equation}
\dot{E}_{\rm GW}= \frac{2}{16\pi} \sum_{\ell =2}^{\ell=8}\sum_{m=1}^{m=\ell} (m\Omega)^2 \, |R \, h_{\ell m}|^2\,,
\end{equation}
where $R$ is the distance to the observer and $h_{\ell m}$ is the harmonically-decomposed gravitational wave metric perturbation. In addition to this, there is also energy flux into any trapped surfaces, black hole horizons, that can be modeled by solving the Teukolsky equation perturbatively in a post-Newtonian expansion (see e.g.~\citep{O28-02_Tanaka:1996ht,O28-02_Mino:1997bw}). Of course, if there are additional sinks of energy, these must also be accounted for here by adding contributions to the total energy flux. 

\subsection{Gravitational Wave Signatures}

In the previous sections, we have briefly summarized some salient features of gravitational wave modeling for comparable-mass and extreme mass-ratio quasi-circular inspirals, but are modifications in this modeling observable?  A plausibly criterion to decide whether a certain gravitational wave modification is detectable by a given detector is the following:
\begin{equation}
\delta \phi_{\rm GW} \geq
\left\{
\begin{array}{lr}
10/\rho & \quad {\rm{if}} \; \rho \geq 10\,,
\\
0  &  \quad  {\rm{if}} \; \rho \leq 10\,,
\end{array}
\right.
\label{O28-02_SNR-measure}
\end{equation}
where $\rho^{2}$ is the square of the signal-to-noise ratio (SNR), defined as
\begin{equation}
\rho^{2}(h) = 4 \int \frac{df}{S_{h}(f)} |\tilde{h}|^{2}\,.
\end{equation}
As before, $\tilde{h}$ is the Fourier transform of the measured gravitational wave response function, while $S_{h}(f)$ is the spectral noise density curve of the detector. The piece-wise nature of this criterion accounts for the fact that there is a threshold SNR below which the signal cannot be detected in the first place. 

The above requirements give us a rough sense of when a certain modification might be detectable, but can such a modification be distinguishable from a vacuum waveform. To address this question, one can consider the SNR of the waveform difference
\begin{equation}
\rho^2(\delta h) \equiv \min_{\lambda_2} \left[4 \int \frac{df}{S_{n}(f)} \left|\tilde{h}_{1}(f) - \tilde{h}_{2}(f;\lambda_2)\right|^2 \right],
\end{equation}
where $\tilde{h}_{1}$ and $\tilde{h}_{2}$ are the Fourier transforms of two waveforms (the ``signal'' and ``template''), normalized such that $\rho(h_1)=\rho(h_2)=1$. The template depends on parameters $\vec{\lambda}_2$ that may be different from the true astrophysical ones, where the minimum difference corresponds to the best fit. One can then consider this SNR, minimized over template parameters, to construct a measure similar to that of Eq.~\eqref{O28-02_SNR-measure}. The minimization will partially account for the effect of possible template parameter degeneracies with non-vacuum or non-General Relativity effects.

\section{Astrophysical Imprints}

In this section, we consider two astrophysical environment effects in the modeling of gravitational waves. We begin with massive perturbers, summarizing the results of~\citep{O28-02_Yunes:2010sm}, and then move on to accretion disk effects, summarizing the results of~\citep{O28-02_Kocsis:2011dr,O28-02_Yunes:2011ws}. I refer the interested reader to these papers and the many references therein, which will not be included in this paper for simplicity. 

\subsection{Massive Perturbers}

\begin{figure*}[ht]
\centering
\includegraphics[keepaspectratio=true,width=7cm]{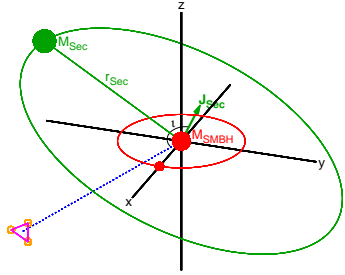}
 \caption{\label{fig:O28-02_sketch} Schematic view of the extreme-mass ratio system (in the $xy$~plane), the massive perturber $M_{\rm Sec}$ (at a distance $r_{\rm Sec}$), and the line of sight.  $\iota$ is the inclination between the primary-secondary supermassive black hole's orbital angular momentum vector and the line of sight.}
\end{figure*}

Let us consider an extreme-mass ratio inspiral, where the stellar-mass compact object is in a quasi-circular, equatorial orbit around a spinning supermassive black hole. Let us now imagine that there is a third compact object, with mass $M_{\rm Sec}$ and at a distance $r_{\rm Sec}$ from the supermassive black hole. We here imagine that $M_{\rm Sec}$ is comparable to $M_{\rm SMBH}$ and much greater than $m_{\rm CO}$. This scenario is depicted in Fig.~\ref{fig:O28-02_sketch}.
Clearly, this is a three-body system, composed of two sub-binaries: the extreme-mass ratio system and the $M_{\rm SMBH}-M_{\rm Sec}$ system. Because of this, the extreme mass-ratio gravitational waves will be emitted from an {\emph{accelerated}} frame (a frame that rotates due to the presence of the secondary). 

The modifications to the trajectories of the small body can be modeled in the framework of the effective-one-body scheme described in Sec.~\ref{sec:O28-02_EOB}. The dominant effect is simply a Doppler shift in the extreme mass-ratio inspiral gravitational wave frequencies, which then leads to an integrated modification in the gravitational wave phase. The implementation of this correction is simple: divide the right-hand-side of Eq.~\eqref{O28-02_EOB-H-Eqs} by the appropriate Doppler factor
\begin{equation}
\dot{\Phi} \equiv \Omega \quad \rightarrow \quad \dot{\Phi} = \Omega \; \left[1 + v_{\rm los}(t,\delta=\pi/2) \right]\,,
\label{O28-02_MP_Eq}
\end{equation}
where $v_{\rm los}$ is the velocity along the line of sight
\begin{equation}
v_{\rm los}(t)=\left( \frac{M_{\rm Sec}}{M_{\rm Tot}}\right)
v_{\rm Newt} \cos\left(\omega_{\rm Newt} t + \delta \right) \; \sin\left(\iota\right)\;.
\label{O28-02_MP_vlos}
\end{equation}
with $v_{\rm Newt} = (G M_{\rm Tot}/r_{\rm Sec})^{1/2}$ the Newtonian virial velocity, 
$\Omega_{\rm Newt} = (G M_{\rm Tot}/r_{\rm Sec}^{3})^{1/2}$ the Newtonian angular velocity
for an object in a circular orbit, and $\delta$ an initial phase offset, with $(\Omega_{\rm Newt} t + \delta)$ the orbital phase of the $M_{\rm SMBH}-M_{\rm Sec}$ system. A constant relative speed can be re-absorbed in a  redefinition of the masses, and thus, it does not lead to a measurable correction. In Eq.~\eqref{O28-02_MP_Eq}, we have ignored the appropriate Lorentz factor $\Gamma$, since $v_{\rm Newt}/c \ll1$, and we have removed the constant velocity drift in $v_{\rm los}$ by choosing $\delta = \pi/2$.

This simple modeling neglects other gravitational wave generation effects. One could incorporate such effects by introducing an external, vectorial force (the product of the the total mass of the system and the time derivative of the velocity of Eq.~\eqref{O28-02_MP_vlos}) to Hamilton's equations. Such a modeling would require a non-adiabatic evolution, which would lead to eccentric and inclined extreme-mass ratio trajectories. The magnitude of this correction, however, is small, as it scales with the tidal force of the perturber on the center of mass of the extreme mass-ratio inspiral, relative to the stellar-mass compact object's acceleration due to the secondary supermassive black hole. Since the tidal force scales as $\sim M_{\rm Sec}/r_{\rm Sec}^{3}$, this effect is suppressed relative to the acceleration by a factor of $r/r_{\rm Sec} \sim 10^{-4}$ for an extreme-mass ratio inspiral with orbital separation $30 M_{\rm SMBH}$ and a primary-secondary binary orbital separation of $0.01$ pc. This ratio would not be small if the perturber were in the near-zone (less than a gravitational wavelength away) from the center of mass of the extreme-mass ratio system. 

We can now estimate what the distance and mass of the secondary black hole must be for its effect on extreme mass-ratio inspiral gravitational to be detectable by a LISA-like mission. The time-dependent Doppler shift described above leads to a correction in the gravitational wave phase of $\Delta\phi_{\rm GW}= {\dot v}_{\rm los} T_{\rm obs}N/(2c)$, where $N$ is the number of radians in the waveform, $T_{\rm obs}$ is the observation time. Let us further define $\epsilon$ as the detectable fractional phase shift: $\epsilon\equiv\Delta\phi_{\rm GW, detect}/N$, a fiducial value of which is $\epsilon=10^{-7}$, or 0.1 radians over $\sim 10^6$ radians for a typical one-year inspiral. 

\begin{figure*}[ht]
\centering
\includegraphics[keepaspectratio=true,width=7cm]{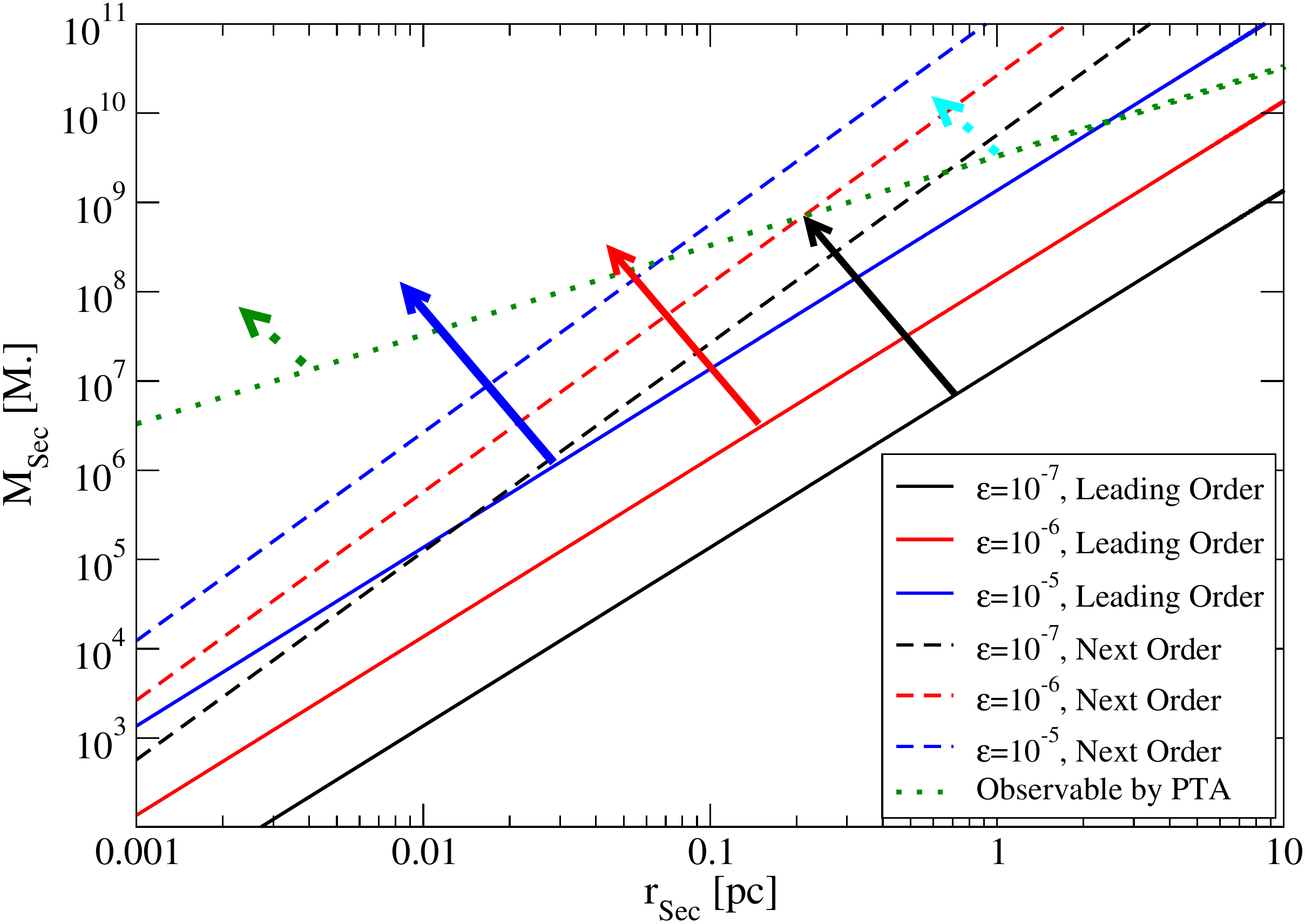}
 \caption{\label{fig:O28-02_FOM} Range of secondary masses and separations that could
 be measurable by LISA given a sufficiently strong EMRI. The region above the solid and dashed
 lines would be observable.  Measurement of the leading-order effect gives a determination
  of the combination $(M_{\rm Sec} \sin\iota)/r_{\rm Sec}^2$, while measuring
  the next-order effect gives a determination of the combination
  $(M_{\rm Sec} \sin\iota)^{3/2}/r_{\rm Sec}^{7/2}$.  Thus measuring both
  effects together allows both $M_{\rm Sec} \sin\iota$ and $r_{\rm Sec}$ to
  be determined.}
\end{figure*}

The definition of $\epsilon$ is a curve in $r_{\rm Sec}-M_{\rm Sec}$ space that delineates the secondary parameters that would lead to a sufficiently large dephasing to be detectable. This is shown in Fig.~\ref{fig:O28-02_FOM} for a range of $\epsilon$ values. Any secondary system with mass and separation above the curves would be observable. The solid curves plot the relation described in the previous paragraph, while the dashed curves account for the next $\ddot{v}_{\rm los} t^{2}$ order effect. For comparison, the region of $(M_{\rm Sec},r_{\rm Sec})$ space that falls in the pulsar-timing-array sensitivity band are also shown. Of course, the gravitational waves that would be detectable by pulsar-timing-arrays would be those generated by the $M_{\rm SMBH}-M_{\rm Sec}$ binary, and their distance to Earth would have to be sufficiently small.
In principle, however, this scenario allows for the possibility of coincident future detection of gravitational waves from LISA-like detectors and pulsar-timing-arrays.

Just because an effect is sufficiently large to be in principle measurable does not imply that it can be distinguished from other vacuum effects. One must therefore worry about possible degeneracies of vacuum system parameters with those introduced by the secondary perturber ($M_{\rm Sec}$ and $r_{\rm Sec}$ in this case). We can get a sense of whether this is the case by computing the Fourier transform of the modified gravitational wave in the stationary-phase approximation. Following Sec.~\ref{sec:O28-02_SPA}, one finds the following correction to the phase
\begin{eqnarray}
\Delta \Psi &=& \frac{3}{128} \left(\pi {\cal{M}} f\right)^{-5/3} \left[ 
\left(\frac{25}{1248} \frac{{\cal{M}} M_{\rm Sec}}{r_{\rm Sec}^{2}}
- \frac{25}{2496} \frac{{\cal{M}} M_{\rm Sec} M_{\rm Tot} t_{c}^{2}}{r_{\rm Sec}^{5}}
\right) \left(\pi {\cal{M}} f\right)^{-8/3}
\right. 
\nonumber \\
&+&\left.
\frac{125}{1032192} \frac{{\cal{M}}^{2} M_{\rm Sec} M_{\rm Tot} t_{c}}{r_{\rm Sec}^{5}}
\left(\pi {\cal{M}} f\right)^{-16/3}
- \frac{625}{1094713344} \frac{{\cal{M}}^{3} M_{\rm Sec} M_{\rm Tot}}{r_{\rm Sec}^{5}}
\left(\pi {\cal{M}} f\right)^{-8}
\right]\,.
\end{eqnarray}
where we have neglected a constant term that is no measurable. Observe that the corrections introduced by $M_{\rm Sec}$ cannot be re-absorbed by a redefinition of vacuum system parameters, which implies that the effect of the secondary are (at worst) weakly-correlated vacuum terms. Moreover, if one could independently constrain the coefficients in from of the dominant $f^{-8/3}$ term and also the $f^{-16/3}$ and $f^{-8}$ terms, one would be able to break the degeneracy between $M_{\rm Sec}$ and $r_{\rm Sec}$ in the leading-order term. 

\subsection{Accretion Disks}

Let us now consider a different astrophysical environment: an extreme mass-ratio system in a circular equatorial orbit that is embedded in the accretion disk associated with the supermassive black hole. We refer the interested reader to~\citep{O28-02_Kocsis:2011dr} for astrophysical scenarios that could lead to this configuration. Of course, the effect of such a disk on the extreme-mass ratio inspiral will depend on the disk properties, which are not accurately known. One must then think of all possible effects that could be present and determine which one leads to the largest correction to extreme-mass ratio inspirals. 

The dominant effect is that induced by the gravitational torque induced by spiral density waves in the accretion disk (ie.~{\emph{migration}} torque in planetary dynamics). Migration can occur whether a gap is cleared out or not, with the former leading to the largest effects. One might naively think that the stellar-object's mass accretion would also greatly modify the object's motion, but this is not the case because of quenching by a variety of processes, predominantly limited gas supply and radiation pressure. Migration then becomes dominant, but its effect on gravitational waves still depends on the disk's properties. If one models the latter as an $\alpha$ (Shakura-Sunyaev) disk (where viscosity is proportional to the total pressure), then all disk effects become negligible. But if the disk is better described by a $\beta$ profile (where viscosity is proportional to gas pressure only) in the radiation-dominated (extreme-mass ratio) regime, then the effect of migration would be measurable. Therefore, if such an accretion disk effect is detected one would know that the disk associated with that particular extreme mass-ratio binary is of $\beta$ type.

\begin{figure*}[ht]
\begin{center}
\begin{tabular}{c}
 \includegraphics[keepaspectratio=true,width=7cm]{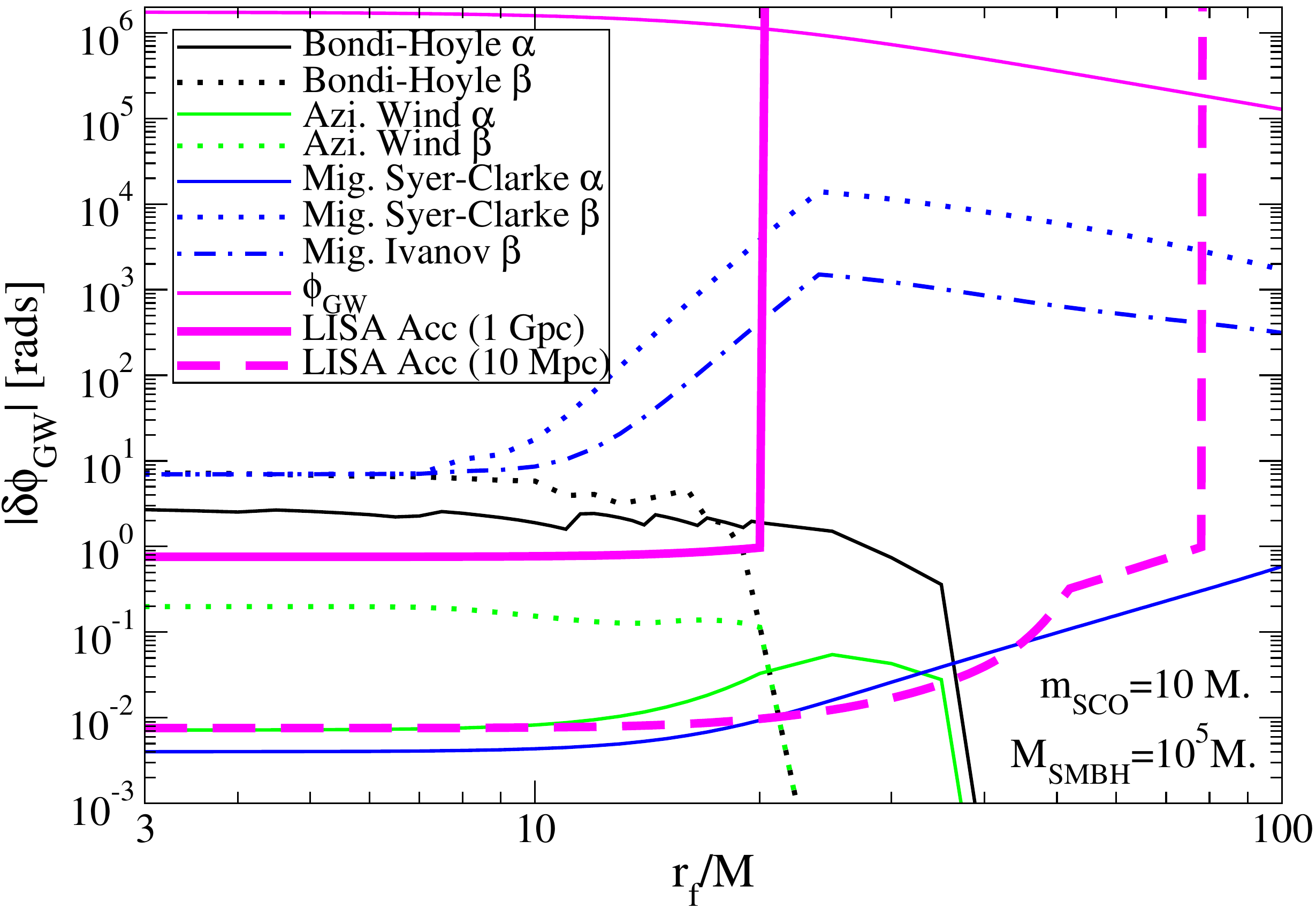}
 \quad
 \includegraphics[keepaspectratio=true,width=7cm]{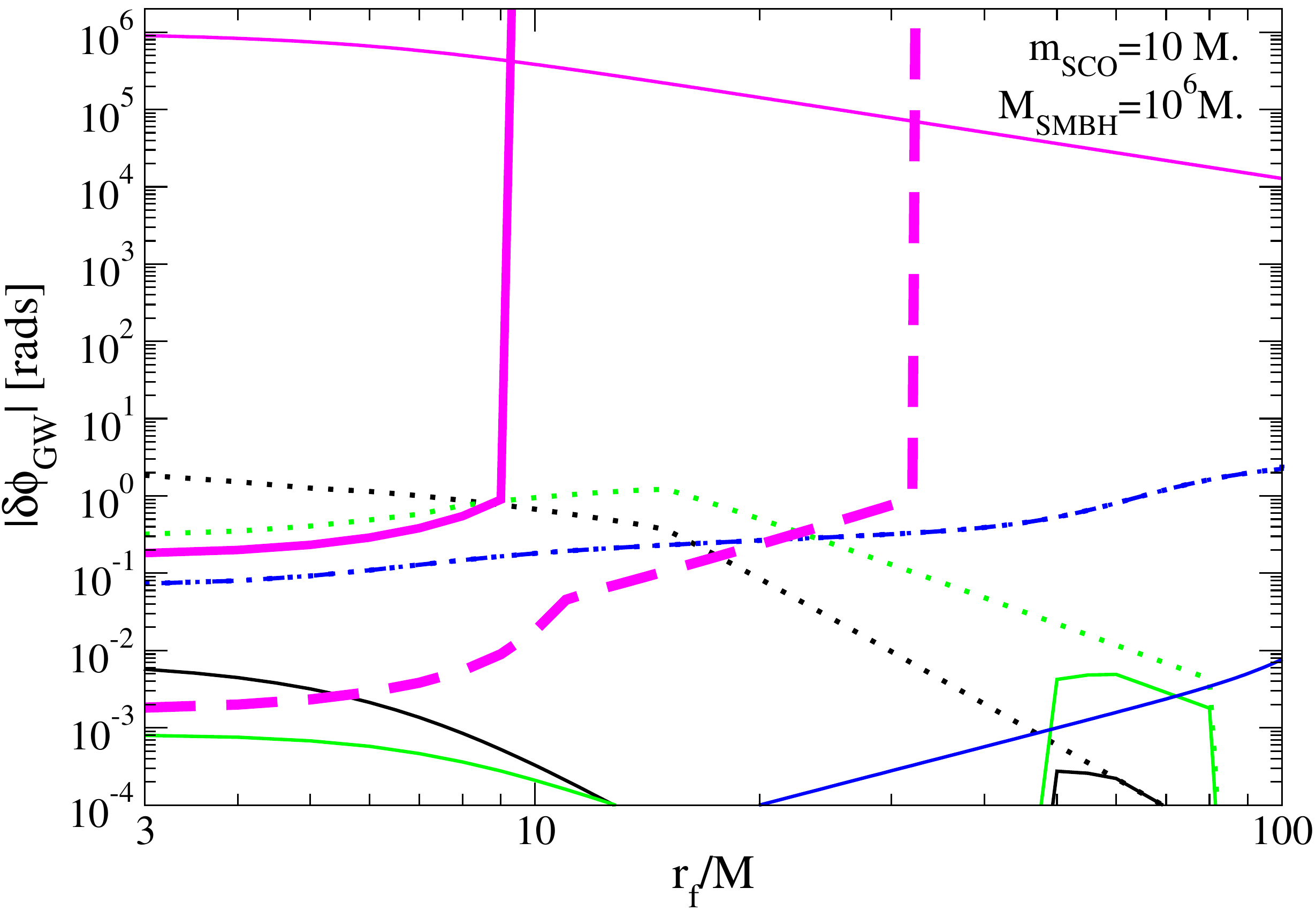}
\end{tabular}
\end{center}
 \caption{\label{fig:O28-02_Disk-FOM} Gravitational wave dephasing as a function of final radius in units of $M_{\rm SMBH}$ induced by different disk effects. Solid (dotted) curves correspond to $\alpha$ ($\beta$) disks, with different colors indicating different disk effects: black corresponds to Bondi-Hoyle-Lyttleton (BHL) accretion onto the small object, green to azimuthal winds and blue to migration. The thin, solid magenta line is the total accumulated gravitational wave phase in vacuum.  The thick, solid (dashed) magenta line corresponds to a measure of the accuracy to which LISA can measure the gravitational wave phase for a source at $1$ Gpc ($10$ Mpc). Observe that certain disk effects, like migration, can leave huge imprints on the gravitational wave observable, inside the LISA accuracy bucket, provided the disk is a $\beta$ one.}
\end{figure*}

Figure~\ref{fig:O28-02_Disk-FOM} summarizes the impact of the main accretion disk effects on the gravitational wave. This figure plots the dephasing (the difference between the gravitational wave phase computed with and without disk effects) as a function of the final orbital radius,  assuming a one year inspiral. The left and right panels show results for different extreme mass-ratios, while the different curve colors and styles correspond to different disk effects: black is for accretion onto the small compact object, blue is for migration and green is for azimuthal winds. The latter corresponds to a difference in the orbital compact object velocity and the gas velocity, which can push the former forward or backward. The total accumulated gravitational wave phase is denoted by a thin magenta line, while the thick magenta lines correspond to two (an optimistic and a pessimistic) measures of the accuracy of a classic LISA-like mission. Notice that migration can lead to dephasing of order $10^{2}-10^{3}$ radians for $\beta$ disks, while the same effect is essentially negligible for $\alpha$ disks. Therefore, a measurement of such disk effects could allow us to infer disk properties related to migration. 

As before, a simple dephasing argument does not guarantee that disk effects can be disentangled from other vacuum system parameters. One can get a sense of whether such disentanglement is possible by computing the Fourier transform of the gravitational wave response function in the stationary-phase approximation:
\begin{equation}
\Psi/\Psi_{\rm vac} = 1 - \tilde{A}_{1} \alpha_{1}^{{c}_{1}} \dot{m}_{\rm SMBH, 1}^{{c}_{2}} M_{\rm SMBH, 5}^{\tilde{a}_{3}} q_{0}^{\tilde{a}_{4}} u_{0}^{\tilde{a}_{5}}\,,
\label{O28-02_disk-phase}
\end{equation}
and
\begin{equation}
|\tilde{h}|/|\tilde{h}|_{\rm vac} = 1 - \tilde{B}_{1} \alpha_{1}^{{c}_{1}} \dot{m}_{\rm SMBH, 1}^{{c}_{2}} M_{\rm SMBH,5}^{\tilde{a}_{3}} q_{0}^{\tilde{a}_{4}} u_{0}^{\tilde{a}_{5}}\,,
\label{O28-02_disk-amp}
\end{equation}
where $\alpha_{1} = \alpha/0.1$ is the normalized $\alpha$-viscosity parameter, $\dot{m}_{\rm SMBH,1} \equiv \dot{m}_{\rm SMBH}/0.1$ is the normalized supermassive black hole accretion rate $\dot{m} \equiv \dot{M}_{\rm SMBH}/\dot{M}_{\rm SMBH, Edd}$ with $\dot{M}_{\rm SMBH, Edd}$ the Eddington rate. Similarly, $M_{\rm SMBH,5} = M_{\rm SMBH}/(10^{5} M_{\odot})$ is a normalized supermassive black hole mass, $q_{0} \equiv q/10^{-4}$ is the normalized mass-ratio $q = M_{\rm SMBH}/m_{\rm CO}$ and $u_{0} \equiv (\pi {\cal{M}} f)/(6.15 \times 10^{-5})$ is a normalized reduced frequency for a gravitational wave frequency of $10^{-2}$ Hz. The parameters $(\tilde{A}_{1},\tilde{B}_{1}, \tilde{a}_{i},c_{i})$ are given in Table~\ref{Table:O28-02_pars2}. Of course, these expressions are only valid when the accretion disk effects are small perturbations away from the vacuum evolution (ie.~at sufficiently small separations).
\begin{table*}[htbp]
\begin{center}
\begin{tabular}{c c c c c c c c c}
 & $\tilde{A}_{1}$ & $\tilde{B}_{1}$ & $\tilde{a}_{3}$ & $\tilde{a}_{4}$ & $\tilde{a}_{5}$ & $c_{1}$ & $c_{2}$ \\
\hline
BHL \, $\alpha$ & $3\,(-8)$ & $2\,(-7)$  & $1$ & $4$ & $-20/3$ & $-1$ & $-5$ \\
BHL \, $\beta$ & $1\,(-5)$ & $1\,(-4)$ & $6/5$ & $79/25$ & $-79/15$ & $-4/5$ & $-17/5$ \\
\hline
W \, $\alpha$ & $6\,(-17)$ & $1\,(-16)$ & $1$ & $16/5$ & $-16/3$ & $-1$ & $-3$ \\
W \, $\beta$ & $6\,(-12)$ & $4\,(-11)$ & $6/5$ & $59/25$ & $-59/15$ & $-4/5$ & $-7/5$ \\
\hline
M1\, $\alpha$ & $3\,(-10)$ & $4\,(-9)$ & $1$ & $16/5$ & $-16/3$ & $-1$ & $-3$ \\
M1\, $\beta$   & $1\,(-6)$ & $3\,(-6)$ & $6/5$ & $59/25$ & $-59/15$ & $-4/5$ & $-7/5$ \\
M2\, $\alpha_{\rm SC}$ & $8\,(-6)$ & $2\,(-5)$ & $1$ & $-2/5$ & $-8/3$ & $0$ & $1$ \\
M2 \, $\beta_{\rm SC}$ & $6\,(-3)$ & $2\,(-2)$ & $1/4$ & $-1/8$ & $-25/12$ & $1/2$ & $5/8$ \\
M2 \, $\beta_{\rm IPP}$ & $6\,(-4)$ & $2\,(-3)$ & $4/7$ & $-17/70$ & $-7/3$ & $2/7$ & $11/4$ \\
\end{tabular}
\caption{\label{Table:O28-02_pars2} Columns are parameters in Eqs.~\eqref{O28-02_disk-phase} and~\eqref{O28-02_disk-amp} and rows are disk effects. The notation $x\,(y) = x
\times 10^{y}$ in radians for $\tilde{A}_{1}$ and dimensionless for
$\tilde{B}_{1}$. Observe that the frequency exponent $\tilde{a}_{5} < 0$, implying
that these accretion disk effects are dominant at small frequencies (large radii).}
\end{center}
\end{table*}

The important result here is that the dependence of the disk effects with frequency, ie.~the parameter $\tilde{a}_{5}$, is of opposite sign with respect to those that arise in the vacuum gravitational wave phase. This is because disk corrections are largest for large radii, equivalent to weak-field vacuum effects, while post-Newtonian corrections to the vacuum gravitational wave phase are largest for small radii, equivalent to strong-field effects. This suggests that migration effects are weakly correlated with vacuum system parameters, as one could not reabsorb these disk effects by choosing different vacuum parameters.

\section{Modified Gravity Imprints}

In this section, we discuss how modified gravity imprints can manifest in the gravitational wave observable. We describe a scheme developed to constrain such imprints, the so-called parameterized post-Einsteinian (ppE) framework. We mainly follow the presentation in~\citep{O28-02_Yunes:2009hc} and the recent results of~\citep{O28-02_Cornish:2011ys}. 

\subsection{Top-Down Approach}

One approach to testing Einstein's theory with gravitational wave observations is a top-down one: pick a modified gravity theory, study its consequences and match-filter gravitational wave data with a template bank specifically constructed for that theory to test for possible deviations consistent with that theory. Most studies carried out so-far considered only quasi-circular coalescences in the inspiral phase. For such waveforms, the modified Fourier transform of the response function in the stationary-phase approximation can be modeled as
\begin{equation}
\tilde{h}(f) = \tilde{h}_{\rm GR}(f) \left(1 + \alpha u^{a}\right) e^{i \beta u^{b}}\,,
\end{equation}
where the GR transform was already described in Sec.~\ref{sec:O28-02_SPA}, while $(a,b)$ are numbers and $(\alpha,\beta)$ are functions of system parameters and parameters of the theory. We present their values in the Table~\ref{Table:O28-02_ppEpars}.
\begin{table*}[htbp]
\begin{center}
\begin{tabular}{c c c c c}
Theory  & $\alpha$ & $a$ & $\beta$ & $b$ \\ 
\hline
Brans-Dicke & $0$ & $\cdot$ & $-\frac{5}{3584} \frac{S^{2}}{\omega_{\rm BD}} \eta^{2/5}$ & $-\frac{7}{3}$ \\ 
Massive Graviton & $0$ & $\cdot$ & $- \frac{\pi^{2} D {\cal{M}}}{\lambda_{g}^{2} (1 + z)}$ & $-1$ \\ 
Parity Violation & $4 v \delta {\cal{A}}$ & $1$ & $0$ & $\cdot$ \\ 
Non-Dynamical CS Gravity & $4 \pi \delta\dot{\vartheta}_{\rm CS}  \delta {\cal{A}}$ & $1$ & $0$ & $\cdot$ \\ 
$G(t)$ Theory & $-\frac{5}{512} \dot{G} {\cal{M}}$ & $-\frac{8}{3} $ & $-\frac{25}{65536} \dot{G}_{c} {\cal{M}}$ & $-\frac{13}{3}$ \\ 
Lorentz Violation & $0$ & $\cdot$ & $-\frac{\pi^{2 - \gamma}}{(1 - \gamma)} \frac{D_{\gamma}}{\lambda_{\rm LV}^{2 - \gamma}} \frac{{\cal{M}}^{1- \gamma}}{(1  + z)^{1- \gamma}}$ & $-\alpha_{\rm LV} - 1$ \\  
Conservative EDGB & $\frac{5}{6} \eta^{-4/5} \zeta_{\rm EDGB}$ &  $\frac{4}{3}$ & $\frac{25}{64} \eta^{-4/5} \zeta_{\rm EDGB}$ & $-\frac{1}{3}$ \\
Extra Dimensions & $0$ & $\cdot$ & $-\frac{75}{2554344} \frac{dM}{dt} \eta^{-4} (3- 26 \eta + 24 \eta^{2} )$ & $-\frac{13}{3}$ \\
\end{tabular}
\caption{\label{Table:O28-02_ppEpars} Parameters that define the deformation of the response function in a variety of modified gravity theories. The notation $\cdot$ means that a value for this parameter is irrelevant, as its amplitude is zero. All other parameters are defined in the text.}
\end{center}
\end{table*}

Let us explain how the symbols of Table~\ref{Table:O28-02_ppEpars} are defined. In Brans-Dicke theory, $S$ is the difference in the square of the sensitivities and $\omega_{\rm BD}$ is the Brans-Dicke coupling parameter~\cite{O28-02_Berti:2005qd}. In the phenomenological massive graviton theory, $D$ is a certain distance measure, $z$ is redshift and $\lambda_{g}$ is the Compton wavelength of the graviton~\cite{O28-02_Berti:2005qd}. In the parity violation case, the amplitude deformation depends on the inclination angle and the beam pattern angles via~\citep{O28-02_Yunes:2010yf}
\begin{equation}
\delta {\cal{A}} = \frac{\left( F_{+}^{2} + F_{\times}^{2} \right) c_{i} \left(1 + c_{i}^{2}\right)}{F_{+}^{2}\left(1 + c_{i}^{2}\right) + 4 F_{\times}^{2} c_{i}^{2}}\,,
\end{equation}
while $v$ is a measure of parity violation. In non-dynamical Chern-Simons (CS) gravity~\citep{O28-02_Alexander:2009tp}, the quantity $\delta \dot{\vartheta}_{\rm CS}$ is the time derivative of the CS scalar field (ie.~the magnitude of the canonical embedding coordinate)~\cite{O28-02_Yunes:2010yf}. In a varying Newton's constant theory, $\dot{G}_{c}$ is the value of the time derivative of Newton's constant at coalescence~\citep{O28-02_Yunes:2009bv}. In the phenomenological Lorentz-violating theory, $\lambda_{\rm LV}$ is a distance scale at which Lorentz-violation becomes important, while $\gamma$ is the graviton momentum exponent in the deformation of the dispersion relation~\citep{O28-02_Mirshekari:2011yq}. In Einstein-Dilaton-Gauss-Bonnet gravity, $\zeta_{\rm EDGB}$ is the coupling parameter in the theory, and we here present only the modification in the response function introduced by the deformation of the background Schwarzschild metric~\citep{O28-02_Yunes:2011we}. In theories with extra dimensions, $dM/dt$ is the mass loss due to gravitons leaking into the extra dimension. 

As one can see from the above table, there are many possible deviations from General Relativity that one could consider, and none of the above has been sufficiently well-studied to determine whether any of them are compelling alternatives. A top-down approach would require us to construct as many template banks as there are alternative theories to consider. Of course, each of these banks would contain additional dimensions in parameter space (and thus many more templates), as one would have to allow for variations of the fundamental parameters of the modified theories. This might increase the computational cost beyond current capabilities. 

But perhaps the biggest drawback of a top-down approach is that one must pick a theory before-hand, and this does not allow the data to select the theory that fits it best. Instead, this choice is made by the gravitational wave theorist or data analyst. It is entirely possible that, if General Relativity is incorrect in the strong field, none of the above theories will represent the correct modification. Thus, if one implements a top-down approach, one is forcing a certain fundamental bias into the analysis, ie.~the fundamental bias that we, as theorists, know what modifications of gravity are possible. 

\subsection{Bottom-Up Approach}

A new scheme has been proposed to alleviate this fundamental bias: the ppE framework~\citep{O28-02_Yunes:2009hc}. In this scheme, one enhances General Relativity templates $\tilde{h}_{\rm GR}$, which depend on system parameters $\vec{\lambda}_{\rm Sys}$, through the introduction of theory parameters $\vec{\lambda}_{\rm ppE}$:
\begin{equation}
\tilde{h}(f;\vec{\lambda}_{\rm Sys}) \rightarrow \tilde{h}_{\rm GR}(f;\vec{\lambda}_{\rm Sys}) + \delta \tilde{h}(f;\vec{\lambda}_{\rm Sys};\vec{\lambda}_{\rm ppE})\,.
\end{equation}
When these parameters acquire certain values, the templates reduce exactly to those predicted in General Relativity, ie.~$ \delta \tilde{h}(f;\vec{\lambda}_{\rm Sys};\vec{\lambda}_{\rm ppE}^{\rm GR})=0$. When they acquire other values, the templates reduce to the predictions of modified gravity theories. The idea then is to match filter or perform Bayesian statistics with this enhanced template bank to allow the data to select the best-fit values of $\vec{\lambda}_{\rm ppE}$ that best fit the signal.

This framework emerges from a generalization of the parameterized post-Newtonian scheme, developed in the 1970's to test modified gravity theories with Solar System experiments~\citep{O28-02_1970ApJ...161.1059N,O28-02_1971ApJ...163..595T,O28-02_1971ApJ...163..611W,O28-02_1971ApJ...169..125W,O28-02_1972ApJ...177..757W,O28-02_1972ApJ...177..775N}. That scheme proposed the enhancement of the main quantity which all Solar System observables depend on, the weak-field expansion of the metric tensor in a certain coordinate system and gauge, by theory (ppN) parameters:  
\begin{equation}
g_{\mu \nu}(x^{\mu};\vec{\lambda}_{\rm Sys}) \rightarrow g_{\mu \nu}(x^{\mu};\vec{\lambda}_{\rm Sys},\vec{\lambda}_{\rm ppN})\,.  
\end{equation}
When these parameters vanish, then the metric tensor reduces exactly to that predicted by General Relativity $g_{\mu \nu}(x^{\mu};\vec{\lambda}_{\rm Sys},0)$, while when they don't, they describe the weak-field expansion of the metric tensor in a plethora of modified gravity theories. The idea is then to constrain the values of $\vec{\lambda}_{\rm ppN}$ by measuring Solar-System observables that depend on the metric tensor. 

The ppN and ppE frameworks, however, differ in the priorities that drive the parameterized General Relativity deviations. While the former does not care about how many theory ppN parameters are introduced, in the latter we wish to minimize the number of ppE parameters. Otherwise, the inclusion of too many ppE parameters dilutes the information in the extracted best-fit, raising the false-alarm probability. One must then find a balance between the number of parameters introduced and the amount of bias contained in the templates.

The precise form of ppE templates clearly depends on the system under consideration and the level of theoretical complexity that one wishes to include. Keeping in line with the rest of this paper, we concentrate here on non-spinning, quasi-circular inspirals. In their original work,~\citep{O28-02_Yunes:2009hc} also proposed ppE templates that contain the merger and ringdown phases in a hybrid fashion, but we will not discuss these here. As for the theoretical complexity, this depends on the number of additional ppE theory parameters one wishes to include. 

The simplest ppE realization, which we refer to as the {\emph{restricted ppE templates}}, is the following~\citep{O28-02_Arun:2006yw,O28-02_Mishra:2010tp}
\begin{equation}
\delta \tilde{h}(f;\vec{\lambda}_{\rm Sys};\alpha,\beta) = \tilde{h}_{\rm GR} \left(\alpha \; u^{a_{\rm PN}} \right) e^{i \beta u^{b_{\rm PN}}}\,,
\end{equation}
where the theory parameters $(\alpha,\beta)$ are to be searched over, while $(a_{\rm PN},b_{\rm PN})$ are numbers equal to some of the post-Newtonian predictions for amplitude and phase frequency exponents, ie.~$(a_{\rm PN},b_{\rm PN})$ are not allowed to vary. For example, a Newtonian prediction for the amplitude and phase frequency exponents would be $a_{\rm Newt} = -7/3$ and $b_{\rm Newt} = -5/3$, while a 1.5PN prediction would be $a_{1.5PN} = 1$ and $b_{1.5PN} = -2/3$. 

One can think of these tests as consistency checks of the post-Newtonian expansion. That is, given a detection of a gravitational-wave, one can in principle measure the chirp mass and the mass ratio by detecting the Newtonian and 1PN corrections to the phase. Allowing for a $1.5$PN parameterized deviation in the phase then allows one to check whether this term is indeed as predicted by the post-Newtonian expansion. Alternatively, one can also think of these tests as straw-men indicators for GR consistency/inconsistency. That is, given a low signal-to-noise ratio detection, one can ask whether the signal detected is consistent with a General Relativity hypothesis $(\alpha,\beta) = (0,0)$~\cite{O28-02_Li:2011cg}.

The great disadvantage of the restricted ppE formalism is that, although one might be able to discern whether deviations are present, one cannot tell what kind of deviations those are. To alleviate this problem, one can introduce {\emph{free ppE templates}}, the simplest version of which is~\citep{O28-02_Yunes:2009hc} 
\begin{equation}
\delta \tilde{h}(f;\vec{\lambda}_{\rm Sys};\alpha,\beta) = \tilde{h}_{\rm GR} \left(\alpha \; u^{a} \right) e^{i \beta u^{b}}\,,
\end{equation}
where now $(\alpha,\beta,a,b)$ are free ppE parameters. Observe that different choices of ppE parameters can recover all the modified theory predictions listed in Table~\ref{Table:O28-02_ppEpars}. Furthermore, one can show that a generic power-series deformation of the binding energy or the gravitational wave luminosity will lead to a metric with the above ppE parameterization~\citep{O28-02_Yunes:2009hc}.

Given a stretch of gravitational wave data, one can then search via matched filtering with free ppE templates and allow the data to select the $(\alpha,\beta,a,b)$ ppE parameters that fit it best. If a deviation is present, such a procedure would not only signal the presence of a deviation, but it would also signal what type of phase or amplitude deformation is introduced. That is, one would know the best-fit numerical values of $a$ and $b$, which encode what type of modified gravity effect is present. For example, as suggested by Table~\ref{Table:O28-02_ppEpars}, a non-zero detection of $\beta$ associated with $b = -7/3$ would signal the presence of a scalar dipolar mode, while one associated with $b = -13/3$ would signal the existence of an anomalous acceleration. 

\begin{figure*}[ht]
\centering
\includegraphics[keepaspectratio=true,width=7cm]{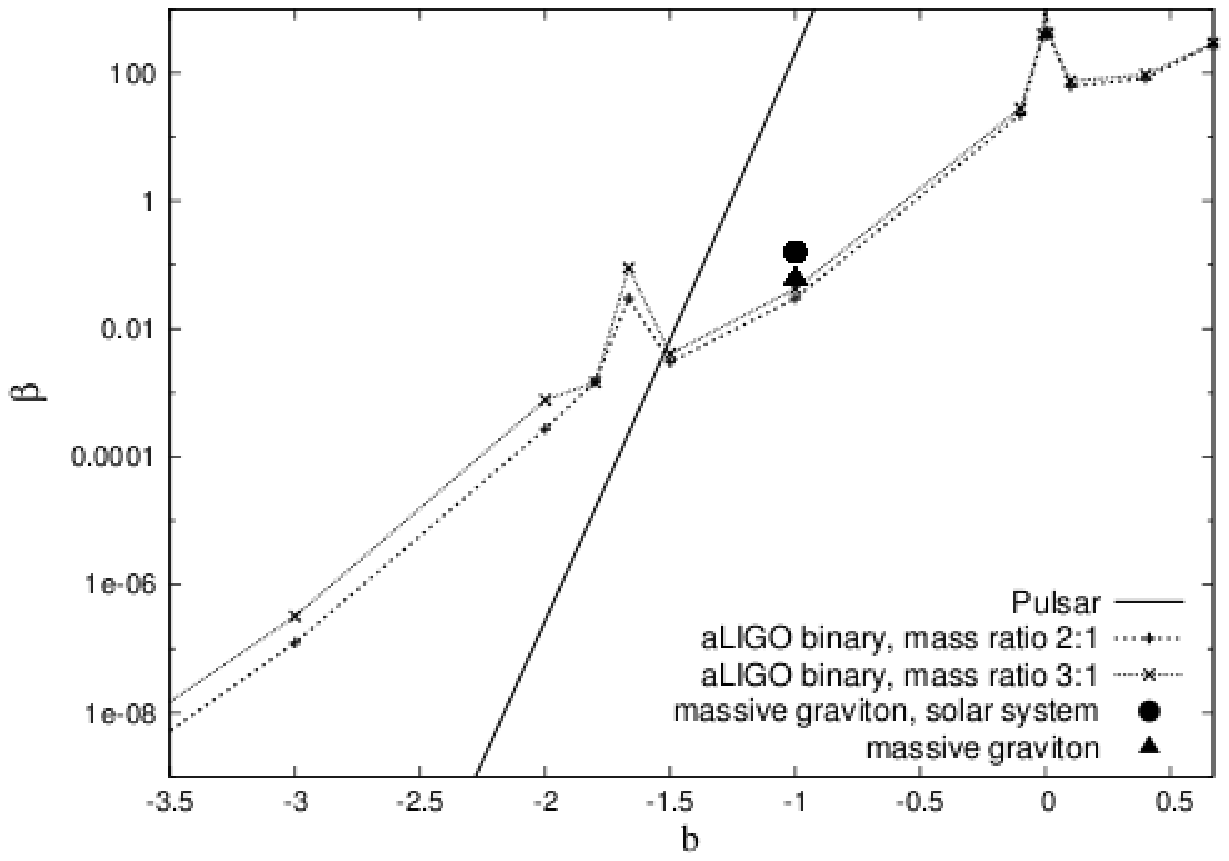}
\includegraphics[keepaspectratio=true,width=7cm]{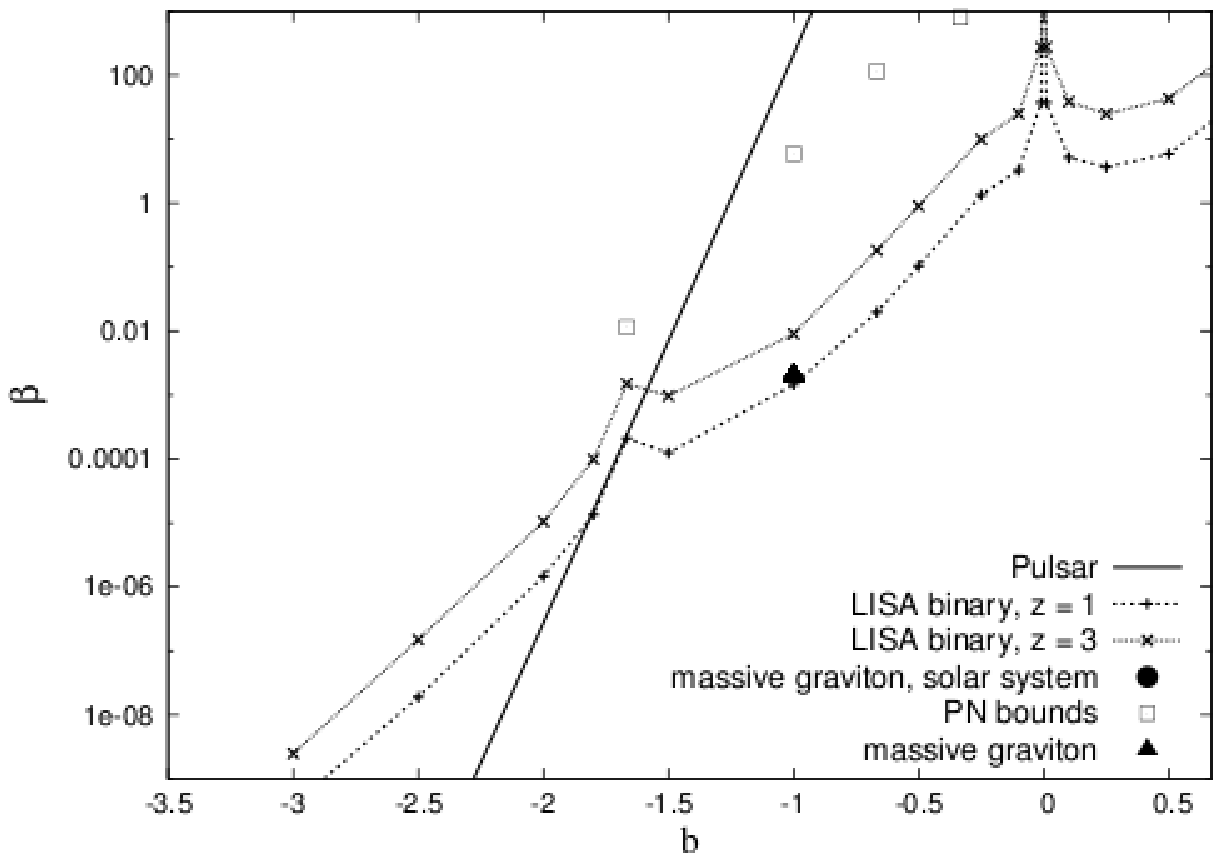}
 \caption{\label{fig:O28-02_beta_bounds} Left: $(3 \sigma)$-bounds on $\beta$ for different values of $b$ for a single ${\rm SNR}=20$ aLIGO/aVirgo detection. The two dotted lines correspond to two sources with different mass ratios, total masses, and sky locations, but both at redshift $z = 0.1$ ($D_{L} = 462 Mpc$).  The solid line is the $(3 \sigma)$-bound on $\beta$ from the golden pulsar (PSR J0737-3039), while other symbols are bounds from Solar System experiments and other aLIGO analyses. Right: Same as left but for two classic-LISA sources at redshift $z=1$ and $z=3$.}
\end{figure*}
We have recently carried out a detailed Bayesian analysis with free ppE waveforms, whose main results are shown in Fig.~\ref{fig:O28-02_beta_bounds}~\citep{O28-02_Cornish:2011ys}. This figure shows the $(3 \sigma)$-bounds on $\beta$ for different values of $b$ (ie.~an exclusion plot for anything above the dotted curves), given gravitational wave detections consistent with General Relativity. The different dotted curves correspond to different binary systems, where we consider the inspiral phase only. For comparison, we also plot current bounds on $\beta$ from the double binary pulsar, as well as a few Solar System constraints. Observe that for $b\gtrsim-2$, gravitational wave observations could rule out a large sector of $(\beta,b)$ parameter space that is currently unconstrained. These results where confirmed using both a Fisher analysis as well as mapping out the likelihood surface with a Markov-Chain Monte-Carlo techniques. 

Given a gravitational wave detection, one can also ask whether a General Relativity waveform (an Einstein hypothesis) or a non-General Relativity waveform (a non-Einstein hypothesis) best fits the data. This can be established by computing the Bayes factor or the odds-ratio of these two hypothesis. When using ppE templates, the hypothesis become nested, but the above described analysis can still be easily carried out. The Bayes factor will increase the larger $(\alpha,\beta)$ are, and so assuming a threshold Bayes factor of $10$ (1 to 10 betting odds that one of the two hypothesis is incorrect and the other is correct) one can derive the accuracy to which $(\alpha,\beta)$ can be measured as a function of $b$. Implementing such a scheme, one obtains similar results to those already plotted in Fig.~\ref{fig:O28-02_beta_bounds}.

\section{A Roadmap for the Future}

The era of multi-messenger gravitational wave astrophysics is at our doorstep. A full exploitation of gravitational wave information will require a deep collaboration between astrophysicists, gravitational wave modelers, general relativists and data analysts. In this paper, I have described recent efforts to extract information about the astrophysical and theoretical environment in which compact binaries might be evolving.      

On the astrophysical side, we have seen that the presence of a secondary perturber or an accretion disk can lead to observable effects in extreme-mass ratio inspiral gravitational waves. As for the former, these effects become non-negligible when secondary perturbers are at a distance of $0.1$ pc or less from the center of mass of the extreme-mass ratio binary, given secondary masses of $10^{7} M_{\odot}$. As for the latter, angular momentum transport due to density waves, analogous to planetary migration, can leave observable signatures in extreme-mass ratio binaries embedded in the thin disks, provided the latter are of $\beta$-type. These effects seem weakly degenerate with other system parameters, as suggested by a simple Fourier analysis. 

On the fundamental theory side, we have seen that a top-down approach leads to a variety of waveform predictions. There is no particular modified gravity theory that is more compelling than General Relativity, which forces us to consider all theories on an equal footing. Performing a data analysis study for all such theories might be computationally prohibitive. Instead, one can carry out a bottom-up approach by match-filtering with ppE templates. These waveforms enhance General Relativity templates with certain well-motivated theory parameters, which then allows the data to select the parameters that best fit it, lifting some degree of fundamental bias.  

A variety of paths present themselves for future studies. One could, for example, consider other astrophysical environment effects, or other specific top-down modified gravity theories, to discover what the corrections to the waveform are. Alternatively, one could consider whether the presence of an astrophysical environment could prohibit tests of Einstein's theory. This seems quite unlikely, given that modified gravity effects if present should be there always, while astrophysical environment effects should only affect a hand-full of sources.  Finally, a realistic implementation of the ppE idea will require the generalization of the free ppE model to more complex and accurate waveforms. If this is not accomplished, mismodeling bias is likely to contaminate ppE studies, hindering the ability of gravitational waves to constrain Einstein's theory. Work along this lines is promising.

\end{document}